\documentstyle[12pt,worldsci]{article}
\begin{document}
\title{SOLAR NEUTRINOS ALMOST INDEPENDENTLY\\
 OF SOLAR MODELS}
\author{V. Castellani$^1$, S. Degl'Innocenti$^2$, G.Fiorentini$^2$ and
B. Ricci$^3$\\
\vspace{0.3cm}
$^1${\em Physics Department, University of Pisa,
 56100 Pisa, Italy\\
and Osservatorio Astronomico di Collurania, 64100 Teramo, Italy}\\
$^2${\em Physics Department, University of Ferrara,
44100 Ferrara, Italy\\
and INFN Sezione di Ferrara, 44100 Ferrara, Italy}\\
$^3${\em Scuola di Dottorato, University of
Padova 35100 Padova, Italy\\
and INFN Sezione di Ferrara, 44100 Ferrara, Italy}}
\maketitle
\begin{abstract}
 Recent solar neutrino results together with the assumption of a
stationary Sun imply severe constraints on the individual components
of the total neutrino flux : $\Phi_{Be} \leq 0.7 \cdot 10^{9} cm^{-2}
s^{-1}, \Phi_{CNO} \leq 0.6 \cdot 10^{9} cm^{-2} s^{-1}$ and $64 \cdot
10^{9} cm^{-2} s^{-1} \leq \Phi_{pp+pep} \leq 65 \cdot 10^{9} cm^{-2}
s^{-1}$ (at 1$ \sigma$ level), the constraint on $\nu_{Be}$ being in
strong disagreement with $\Phi_{Be}^{SSM} = 5 \cdot 10^{9} cm^{-2}
s^{-1}$.  We study a large variety of non-standard solar models with
low inner temperatures, finding that the temperature profiles T(m)
follow the homology relationship: T(m)=k$T(m)^{SSM}$, so that they are
specified just by the central temperature $T_{c}$.  There is no value
of $T_{c}$ which can account for all the available experimental
results and also if we restrict to consider just Gallium and
Kamiokande results the fit is poor.  Finally we discuss what can be
learned from new generation experiments, planned for the detection of
monochromatic solar neutrinos, about the properties of neutrinos and
of the Sun\footnote{To appear in ``Solar Modeling Workshop'', Seattle,
March 20-24, 1994}.
\end{abstract}
\section{Introduction}
\label{sec-introduction}

The latest results of the Gallium experiments\cite{Gall}$^{,}$
\cite {sage} together with the information arising from the Chlorine
\cite{clorine} and Kamioka \cite{kamioka} experiments and -most
important - with the assumption of a quasi-stationary Sun, yield
severe constraints on the individual components of the solar neutrino
flux, in the hypothesis of standard (zero mass, no mixing, no magnetic
moment...) neutrinos.  We remark that the arguments leading to these
constraints, already outlined in a previous paper \cite{future}, are
essentially independent of solar models.\\[0.2cm] For standard
neutrinos, the nuclear energy production chain has to be extremely
shifted towards the ppI termination so that the abundance of $^{7}Be$
nuclei has to be much smaller than in the SSM; the solar neutrino
problem is thus at the level of $^{7}Be$-nuclei production (and no
more just a matter of the rare $^{8}B$-neutrinos).\\[0.2cm] And what
about non-standard solar models? In this spirit, we analyze several
models, built so as to enhance the ppI termination by lowering the
internal temperature. As well known, another way to shift the nuclear
fusion chain towards the ppI termination -without lowering the
temperature- can be found in the realm of nuclear physics (for example
the hypothesis of a resonance in the $^{3}$He+$^{3}$He channel first
advocated by Fowler \cite{fowler}, see also
\cite{noiprimo}$^{,}$
\cite{noineutrinos}) but in this paper we concentrate on non-standard solar
models
with different internal temperature.\\[0.2cm] As a common feature of
these models we find a homology relation for the temperature profiles,
T(m)=k$T(m)^{SSM}$ where k depends on the parameter which is varied,
but is constant with m=M/M$_{\odot}$ ; in other words, a variation of
the solar temperature in the centre implies a definite variation in
all the inner radiative zone.\\[0.2cm] As a consequence, one clearly
understands how the different components of the neutrino flux depend
(just) on the central temperature $T_{c}$ and, since just $T_{c}$
matters, one can perform a $ \chi^{2} (T_{c})$ analysis comparing the
experimental data with the values of non-standard solar
models.\\[0.2cm] New generation experiments are being planned for the
detection of monochromatic solar neutrinos produced in electron
capture ($^{7}Be + e^{-} \rightarrow ^{7}Li + \nu$) and in the pep
($p+ e^{-} + p , d + \nu$) reactions \cite {arpesella}$^{,}$
\cite{alessandrello}$^{,}$
\cite {raghavan}.  Furthermore, Bahcall \cite{spettro} pointed out that the
Doppler
 effects on  monochromatic neutrinos lines can be used to infer inner solar
 temperatures.  In relation with the foregoing analysis, we discuss what can be
learned from
such future measurements about the properties of neutrinos and of the
Sun.\\[0.2cm]
The results presented in this paper will be discussed more extensively in a
forthcoming
publication \cite{noineutrinos}.

\section{Constraints on solar neutrino fluxes, (almost) independently of solar
models}
\label{sec-constraints}

This section is an update of the constraints \cite{future}$^{,}$
\cite{noineutrinos} on solar neutrino fluxes in the light of the
recent results of Gallex and Sage; therefore here we recall just the
main points and we refer to references 5 and 8 for details. \\[0.2cm]
i)\hspace{0.1cm} We assume, as a working hypothesis, standard
neutrinos and quasi-stationary Sun, that is the solar luminosity is
actually balanced by production of nuclear energy.  In this case the
main components of the solar neutrino flux ($\Phi_{pp+pep} ,
\Phi_{Be}$ and $\Phi_{CNO}$) are constrained by the solar constant:
\begin{equation}
K = \sum_{i} (Q/2 - <E>_{i} ) \Phi_{i}
\end{equation}
where K is the solar constant, Q is the energy released in the fusion
reaction $4p + 2e^{-} \rightarrow \alpha + 2 \nu$ and $<E>_{i}$ is the
average neutrino energy. (We assume that other contributions,
particularly $\Phi_{B}$, can be neglected in eq. (1)).\\[0.2cm]
ii)\hspace{0.1cm} In order to calculate $<E>_{i}$ we take the ratio
$\xi = \Phi_{pep}/ \Phi_{pp+pep}$ from the SSM ($\xi = 2.38 \cdot
10^{-3}$) and similarly the ratio $\eta= \Phi_{N}/ \Phi_{CNO}$ =
0.54.\\[0.2cm] iii)\hspace{0.1cm} The signal $S_{x}$ of any experiment
can be represented as
\begin{equation}
\ S_{x} = \sum_{i} X_{i} \Phi_{i}
\end{equation}
where the weighting factors $X_{i}$ specifies the ``response
function'' (=cross section) of the detector for a unitary flux of
electron neutrinos with energy spectrum as given by the {\em i}th
branch of the nuclear reaction chain. (Clearly the $X_{i}$ are ordered
according to the neutrino energy).\\[0.2cm] iv)\hspace{0.1cm} Again as
a working hypothesis, we assume that the solar model is unable to
estimate the Boron neutrino flux and this flux must be taken from
experiments: Kamiokande or Chlorine (as well known, a choice is needed
because the two experiments give almost conflicting results
\cite{noiprimo}$^{,}$ \cite{noineutrinos}).\\[0.2cm] v)\hspace{0.1cm}
Concerning experiments, we use the weighted average between the Gallex
\cite{Gall} and Sage \cite{sage} results, statistical and systematical
errors being added in quadrature:
\begin{equation}
S_{Ga}=(78 \pm 10) \; SNU
\end{equation}
For the Chlorine experiment \cite{clorine} we take the average of the 1970-1992
results:
\begin{equation}
S_{Cl}=(2.32 \pm 0.23)\; SNU
\end{equation}
whereas the Kamiokande \cite{kamioka} results is:
\begin{equation}
\Phi_{B}^{Ka} = (2.9 \pm 0.42) \; 10^{6}cm^{-2}s^{-1}
\end{equation}
Thus we have four unknowns ( $\Phi_{pp+pep} , \Phi_{Be}$ ,
$\Phi_{CNO}$ and $\Phi_{B}$) and three equations (1),(3) and
alternatively (4) or (5). Choosing the Chlorine result, which allows a
larger interval for the unknowns, we obtain the following bounds on
the neutrino fluxes\cite{future}$^{,}$\cite{noineutrinos} (in unit of
$10^{9}$ cm$^{-2}$ s$^{-1}$):\\

\hspace{1.2cm}$64\leq \Phi_{pp+pep} \leq 65 \hspace{1cm} \Phi_{Be} \leq 0.7
\hspace{1cm} \Phi_{CNO} \leq 0.6 \hspace{1cm}$ at  1$\sigma$
\begin{equation}
 61\leq \Phi_{pp+pep} \leq 65 \hspace{1cm}
 \Phi_{Be} \leq 4.2 \hspace{1cm} \Phi_{CNO} \leq 3.6 \hspace{1.15 cm} {\rm at}
\hspace{0.1cm} 3\sigma
\end{equation}
 In brief, the Gallium result together with the observed solar
luminosity implies that almost all neutrinos, if standard, are from
the ppI termination.  The bounds of eq. (6) are very strict since even
a small flux of other (=more energetic than pp) neutrinos give an
appreciable contribution to the Gallium signal.  This is why an
experimental result with 10\% accuracy can fix the $\Phi_{pp+pep}$ at
the level of about 2\%.  (Note that to derive the bound of eq. (6) the
values chosen for $\xi$ and $\eta$ are unessential, see
\cite{future}).\\[0.2cm] The relevance of these bounds becomes clear
in table 1, where we compare the theoretical estimates for the
neutrinos fluxes of different standard solar models with the
experimental bounds shown in eq. (6). One can see that the results of
different updated standard solar models agree well each other.  It is
remarkable that now the comparison between theory and experiment can
be performed at the level of the individual fluxes and not only at the
level of the total signal.
\begin{table}[t]
\vspace{8.8cm}
\caption{ Comparison of the results
for the predicted flux ($cm^{-2} s^{-1}$) of neutrinos of our updated SSM
with Livermore opacities (CDF94)\cite{future} with the results from
a few recent solar models by other authors and with the experimental
bounds of eq.(6):\\
 BP=Bahcall \& Pinsonneault 1992 ``best model
with diffusion'', table VIII of \cite{pinso}.
\\ TCL=
Turck-Chieze and Lopez 1993 \cite{TCL} ``IS Cpp Recent CNO'' i.e. the
model with $\underline{I}$ntermediate $\underline{S}$creening,
$\underline{pp}$ reaction rate according to \cite{carlson}, meteoric
composition from \cite{AG} with the recent CNO abundance \cite{carbonio}$^{,}$
\cite{azoto}$^{,}$ \cite{ossigeno}, fourth column of
table 5b.\\ Note that the bounds on $^{8}B$ flux arises from the
Kamioka experiment directly.  We also report the capture rates for the
$^{37}Cl$ detector (Hom) and for the Gallium experiments (Gall) in
units of SNU (1 SNU=$10^{-36}$ captures per atom $s^{-1}$).  The
central temperature ($T_{c}$) and the original composition of the
models are also indicated.}
\end{table}
Note that the $\Phi_{Be}$ (that is the $^{7}Be$-nuclei production) has
to be suppressed by a factor seven (at 1$\sigma$) with respect to the
estimate of the SSM. Therefore the problem is now at the level of the
branching between the ppI and ppII termination and playing only on
$^{7}Be + p \rightarrow ^{8}B + \gamma$ does not work.
\vspace{0.1cm}

\section{Non-standard solar models with low internal  temperature}
\label{sec-non}

 In this section we analyze the characteristics of several (low inner
temperature) non-standard solar models, built so as to enhance the
pp-I termination in various ways:\\ i)\hspace{0.1cm} reduction of the
metal fraction Z/X\\ ii)\hspace{0.1cm} reduction (by an overall
multiplicative factor) of the opacity tables\\ iii)\hspace{0.1cm}
increase of the astrophysical factor $S_{pp}$ of the $p + p
\rightarrow d+ e^{+} + \nu$ reaction\\ iv)\hspace{0.1cm} reduction of
the Sun age.\\ All the variations have been performed well beyond the
uncertainties of the standard solar model, i.e. we have really built
non standard solar models.\\[0.2cm] As shown in Figs. 1 a-d , the main
feature of all these models is an homology relation for the inner
temperature profiles:
\begin{equation}
T(m)=kT^{SSM}(m)
\end{equation}
where $m=M/M_{\odot}$ is a mass coordinate and k depends on the
parameter which is varied but does not depend on m.  We have verified
that eq. (7) holds with an accuracy better than 1\% in the internal
radiative zone ($M/M_{\odot} \leq 0.97$ or $R/R_{\odot} \leq$ about
$0.7$) for all the models we consider, but for huge (and really
unbelievable) variations of the solar age.\\[0.2cm] Thus, for each
model the temperature profile is uniquely specified by a scale factor,
that can be taken as the central temperature $T_{c}$. Fig. 2 shows, as
a comparison, the temperature profile for our SSM; one can see that
the temperature decreases rapidly as the mass increases.  Regarding
other physical parameters (pressure P, radius R, and density $\rho$,)
the homology is not a common feature of all the models, that is the
homology for all the physical parameters R, P, T and $\rho$ that we
have already found for variations of Spp \cite{noispp}, is not
verified in particular for change of age or Z/X.\\[0.2cm]
\begin{figure}[p]
\vspace{20cm}
\caption { a-d. The temperature profiles T(m) normalized to
$T^{SSM}$(m) for several non-standard solar models, obtained by
varying different parameters: $spp=Spp/Spp^{SSM}$ (Fig. 1a) , $zx=
(Z/X)/(Z/X)^{SSM}$ (Fig
1b), opa=opacity/opacity$^{SSM}$ (Fig 1c), t=age/age$^{SSM}$ (Fig 1d).
We remind that the estimated ($ 1 \sigma$) relative errors for the values of
these
 parameters used in SSM calculation are:
$\Delta Spp/Spp = 1\%$ ,$\Delta (Z/X)/(Z/X)=10\%$, $\Delta opa/opa \approx
10\%$ ,
$\Delta age/age = 3\%$}
\end{figure}
\begin{figure}
\vspace{7.8cm}
\caption{The temperature profile (normalized to the center value) of our SSM}
\end{figure}
Figure 3 shows the behaviour of the main components to the
experimental signals ($\Phi_{pp}, \Phi_{Be}, \Phi_{B}$) as a function
of the central temperature when varying the input parameters of solar
models.  One can see that no matter how $T_{c}$ is varied, the result
is approximately the same.  The explanation of this behaviour is given
by the homology of the temperature.  The neutrino fluxes are strongly
dependent (through the Gamow factors) on the values of the temperature
in the production regions $T_{i}$ and as usual can be locally
approximated by power laws:
\begin{equation}
\Phi_{i} = c_{i} T_{i}^{\beta_{i}}.
\end{equation}
The homology relationship implies $T_{i}=(T_{c}/T_{c}^{SSM}) \cdot
 T_{i}^{SSM}$ and consequently
\begin{equation}
\Phi_{i} = \Phi_{i}^{SSM}  \cdot (T_{c}/T_{c}^{SSM})^{ \beta_{i}}
\end{equation}
this means that each flux is mainly determined by the central
temperature, almost independently on the way the temperature variation
was obtained.\\[0.2cm] We have also verified that the power laws which
link the variation of the neutrino fluxes with the central temperature
change ($ \Phi_{pp} \propto T_{c}^{-0.6} ,
\Phi_{Be} \propto T_{c}^{9}, \Phi_{B} \propto T_{c}^{21}$) are valid also for
huge
variations of the input parameters (see \cite{noineutrinos} for more
details).\\[0.2cm] The dependence on the temperature being generally
stronger for the $^{8}B$ flux than for the $^{7}$Be flux,by lowering
the temperature the Boron neutrinos are suppressed more strongly than
Berillium neutrinos so that the agreement between these non-standard\\

{\noindent solar models and the experimental results cannot be good
 \cite{noispp}$^{,}$\cite{hata}.}  However one can perform a
 $\chi^{2}(T_{c})$ analysis comparing the experimental data with the
 values of non-standard solar models (we remind that one can do this
 because the different component of the neutrino flux depend (just) on
 the central temperature $T_{c}$).  In this way \cite{noineutrinos}
 one finds that there is no value of $T_{c}$ which can account for all
 the available experimental results; This is partly due to the well
 known ``inconsistency'' between Kamiokande and Chlorine.  It is most
 important to note that, even if we restrict to consider just Gallium
 and Kamiokande results, the fit is poor. The reason is that if one
 tries to reduce $\Phi_{Be}$ in accordance with Gallium data, then
 $\Phi_{B}$ becomes too small in comparison with Kamiokande
 result.\\[0.2cm] As a consequence, if one lowers the $^{7}Be + p
 \rightarrow ^{8}B + \gamma$ cross section (as suggested from recent
 theoretical investigations \cite{langanke} and from the analysis of
 recent data on the Coulomb dissociation of Boron \cite{coulomb}) the
 situation gets even worse, see \cite{noineutrinos}.
\begin{figure}[t]
\vspace{9cm}
\caption{ The behaviour of $\Phi_{pp}$, $ \Phi{Be}$ and $\Phi_{B}$ as a
function of the central temperature
when varying the input parameters of the solar models}
\end{figure}
\section{Future detection of monochromatic solar neutrinos}
\label{sec-future}

In the following we discuss
what can be learnt from a  measurement of the intensity of $^{7}Be$ and pep
neutrinos as planned in  new generation experiments
\cite{arpesella}$^{,}$ \cite{alessandrello}$^{,}$\cite{raghavan}.\\[0.2cm]
Concerning the intensity of the $^{7}$Be-line, we recall the bounds of eqs (6):
at 1$\sigma$ (3$\sigma$) the neutrino flux has to be smaller than 0.7 (4.2)
$\cdot 10^{9} cm^{-2} s^{-1}$ or neutrinos are non-standard.  We recall hovewer
that
also the MSW solution requires a low value for berillium neutrinos.\\[0.2cm]
The pep neutrinos are a good indicator of $\Phi_{pp}$, since the ratio
$\Phi_{pep} /
\Phi_{pp}$ is rather stable.  Fig. 4 shows the value of $\Phi_{pep}$ as a
function
of $\Phi_{Be}$ for several non-standard solar models as to reproduce the
Gallium result
within $3\sigma$.  One can see that even for huge \\

{\noindent variations of the parameters, $\Phi_{pep}$ lies in the
range: $0.1 < \Phi_{pep}[10^{9} cm^{-2} s^{-1}] < 0.2$, whereas the
MSW solution requires $\Phi_{pep} \leq 3 \cdot 10^{7} cm^{-2}
s^{-1}$.}  Thus, measurement of the pep-line intensity will clearly
distinguish non-standard solar models from non-standard
neutrinos.\\[0.2cm]
\begin{figure}[p]
\vspace{6cm}
\caption{ The pep neutrino flux vs. the $^{7}Be$ neutrino one:\\
-for the standard solar model (o) - for several non-standard solar
models adjusted so as to reproduce the Gallex result within $3 \sigma$
(the Boron contribution anyhow being taken from Kamiokande
experiments).  The numbers close to each point represent the
corresponding value of $x/x_{SSM}$.\\ -The values for the MSW
solution, for the best fit (X) and for the 90\% C.L. region (dots),
see also \cite{future}.}
\end{figure}
\begin{figure}[p]
\vspace{6cm}
\caption{ Relation between the temperature $T_{i}$ at the Be-peak (pep-peak)
production and the
central temperature in non standard models, from the numerical
calculation symbols, and from the homology relation
$T_{i} = T_{c} (T_{i}^{SSM}/T_{c}^{SSM})$}
\end{figure}
In a recent paper Bahcall \cite{spettro} pointed out that the shift of
``monochromatic'' (Be and pep) neutrino lines due to Doppler effect
can be measured in foreseeable bolometric neutrino detectors.  Thus
one can infer the solar temperature in the production regions (we
remind that neutrinos from different reactions are produced in
different regions of the Sun). We remark that if one measures the
temperature at one place, by using the homology relations, one knows
it elsewhere, for example, in the center of the Sun.\\[0.2cm] On the
other hand, the homology relation itself is testable - in principle-
by comparing the temperature at two different places, as can be done
by looking the shapes of both the $\nu_{Be}$ and $\nu_{pep}$ lines
(see fig. 5).  This would be a test of the mechanism for energy
transport through the inner Sun, see also \cite{noineutrinos}. This
possibility (although remote) looks to us extremely fascinating.
\section{Conclusions}
\label{sec-Conclusions}
\noindent
i) \hspace{0.1cm} The Gallium result together with the luminosity
constraints implies that almost all neutrinos, if standard, are from
the pp-I termination, see eqs. (6).  Thus the problem is now at the
berillium production level and playing only on the $^{7}Be + p
\rightarrow ^{8}B + \gamma$ cross section does not work. \\[0.2 cm]
ii) \hspace{0.1cm} In a large variety of non-standard solar models,
characterized by a radiative energy transport in the inner
($m=M/M_{\odot} < 0.97$) Sun, the temperature profile T(m) is
homologous to that of the standard Solar Model, see Figs. 1.\\[0.2cm]
iii) \hspace{0.1cm} As a consequence of the homology relations, in all
of these non-standard models the main component of the neutrino flux
depend mainly on the central solar temperature $T_{c}$, see
Fig.3.\\[0.2cm] iv) \hspace{0.1cm} Lowering the central temperature
cannot be a solution of the solar neutrino problem even if we restrict
to consider just Gallium and Kamiokande results.\\[0.1cm] v)
\hspace{0.1cm} Measurement of the $^{7}Be$ and, particularly, of the
pep line intensities will be crucial for deciding about neutrino
properties.\\[0.2cm] vi) \hspace{0.1cm} In principle the homology
relation can be tested in future experiments aimed at the measurements
of inner solar temperatures by looking at the deformation/shift of the
pep and Be line shapes.  This would provide a clear test about the
mechanism of energy transport in the solar interior.\\[0.2cm] {\bf
Acnowledgements}\\ It is a pleasure to thank M. Lissia, J. Bahcall and
A. Smirnov for fruitful discussions.  S. Degl'Innocenti thanks the
Institute for Nuclear Theory at the University of Washington for its
hospitality and the Department of Energy for partial support during
the completion of this work.

\end{document}